\documentclass[%
 reprint,
 amsmath,amssymb,
 aps,
pra,
]{revtex4-2}

\usepackage{graphicx}
\usepackage{dcolumn}
\usepackage{bm}

\usepackage{braket}
\usepackage{orcidlink}
\usepackage{comment}
\begin{document}

\preprint{APS/123-QED}

\title{Secure quantum-enhanced measurements on a network of sensors}

\author{Sean William Moore \orcidlink{0000-0003-0355-2073}}
\email{S.W.Moore@sussex.ac.uk}
\author{Jacob A. Dunningham \orcidlink{0000-0001-7976-3278}}%
 \email{J.Dunningham@sussex.ac.uk}
\affiliation{Department of Physics and Astronomy, University of Sussex, Brighton BN1 9QH, United Kingdom}%


\begin{abstract}
Two-party secure quantum remote sensing (SQRS) protocols enable quantum-enhanced measurements at remote locations with guaranteed security against eavesdroppers. This idea can be scaled up to networks of nodes where one party can directly measure functions of parameters at the different nodes using entangled states. However, the security on such networks decreases exponentially with the number of nodes. Here we show how this problem can be overcome in a hybrid protocol that utilises both entangled and separable states to achieve quantum-enhanced measurement precision and security on networks of any size.
\end{abstract}

\maketitle


\section{Introduction}

Secure quantum remote sensing (SQRS) combines quantum metrology with quantum communications to enable high-precision measurements at a remote location with guaranteed security. It is an interesting paradigm, in part, because it provides a new example of a hybrid quantum technology – one that combines two traditional quantum technologies in a single protocol without having to implement each separately. It also has practical advantages, for example the two parties never have to reveal their measurement basis making the protocol more secure to photon splitting attacks than other secure communication protocols \cite{Moore2023} and security is preserved even if remote parties themselves are not secured. An added advantage is that all of the data processing at the site that is intended to gain the information.

There is a growing body of SQRS protocols in the literature with different approaches. Some distribute entangled states such as Bell states between one party (Alice) and a remote site (Bob)~\cite{Giovanetti2002a,Takeuchi2019a,Yin2020a}. Alice makes a projective measurement on her part of the entangled state and Bob uses  his part to measure a local parameter. He then communicates his measurement result to Alice who can use both sets of results (hers and Bob's) to estimate the parameter. If the remote site is trusted and an eavesdropper does not have access to the quantum communication channel this ensures information privacy. Security can also be checked by tomography on the state at Bob's end to ensure it has not been altered in transit by the actions of an eavesdropper~\cite{Yin2020a}.

Other SQRS protocols achieve security by Alice sending orthogonal separable states with equal probability and ensuring that the identity of each individual state remains secret~\cite{Xie2018a,Huang2019a,Moore2023}. Such an approach avoids the need for entangled states or complicated tomographic measurements to protect against man in the middle attacks on the quantum communication channels. In these protocols Alice sends quantum states to Bob, who chooses at random whether or not to apply his (unknown) phase to the states before measuring them. When no phase is applied, Alice can determine the fidelity of the states at Bob's end. Any deviation from her expected result indicates the possible presence of an eavesdropper who has attacked the quantum channel. For example, in~\cite{Moore2023} Alice sends the four Pauli-X and Pauli-Y eigenstates $\{ |X_+\rangle, |X_-\rangle, |Y_+\rangle, Y_-\rangle\}$ with equal probability to Bob who chooses at random to apply his unknown phase to the probe or not. He sends the results to Alice who, because she knows the initial states, can estimate the unknown phase or check the fidelity of the state providing security against eavesdropper  attacks in a similar way to the BB84 quantum key distribution protocol~\cite{Bennett2014QuantumCP}.

A natural extension of SQRS is to consider networks of sensors~\cite{Komar2014a,Kasai2022a,Shettell2022b}. A global measurement strategy like entangling states over the network nodes can provide quantum enhanced measurement precision for functions of parameters compared to a local estimation strategy~\cite{Knott2016,proctor2017networked,Proctor2018,Rubio_2020b,Eldredge2018a,Ge2018a,Qian2019a,Qian2021a,Bringewatt2021a,Bringewatt2024a}. The greatest advantage is for a sum of the parameters $\phi_b$ held by each Bob, i.e. $\theta = \sum_{b=1}^{N_B} \phi_b$ and that is the function we consider in this paper. In this case entanglement can increase the precision, relative to combining the results of separate measurements, by a factor of $\sqrt{N_B}$, where $N_B$ is the number of Bobs ~\cite{Proctor2018}. The problem, however, is that entangled states can also make it exponentially more difficult to detect an  eavesdropper~\cite{Huang2019a}. The reason for this is that all the Bobs must independently and randomly choose to either measure their parameter or do a fidelity check -- this ensures that Eve cannot attack only the quantum states that will not be used for fidelity checking. However, Alice will only detect Eve if all the Bobs simultaneously choose to perform a fidelity check~\cite{Huang2019a}. Such an occurrence becomes exponentially unlikely as the number of Bobs increases. 
 This problem could be resolved by using secure communications between all the Bobs so they can decide in advance when they should all do a fidelity check. However that introduces additional communication channels adding complexity to the protocol as well as more opportunities for eavesdroppers to intercept information.

\begin{figure}
    \centering
    \includegraphics[width= .48\textwidth]{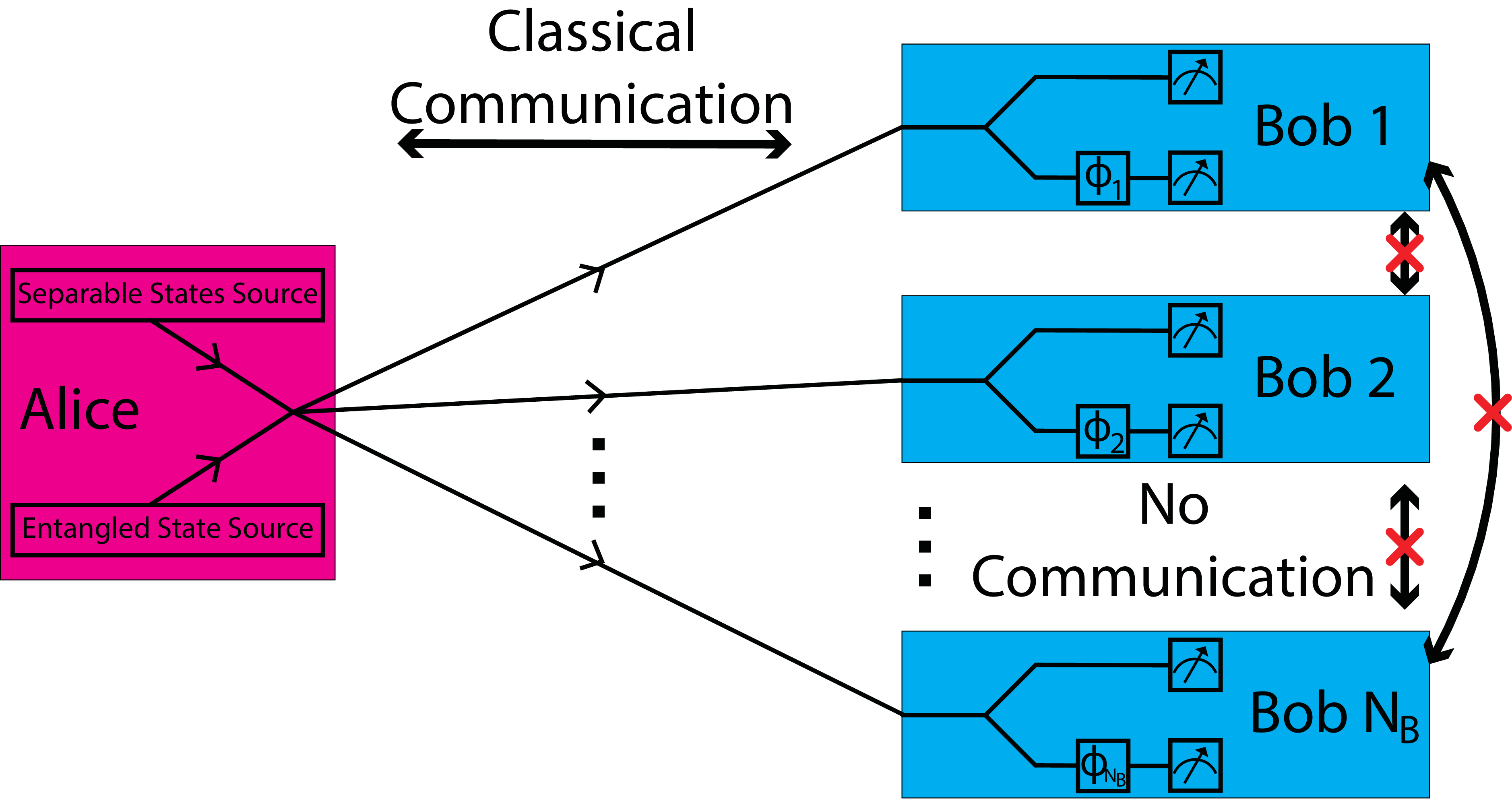}
    \caption{A protocol for estimating functions of quantum parameters held at multiple remote sites while maintaining security from eavesdropping and spoofing. In this scenario each Bob has a parameter and Alice is attempting to estimate some linear function of these parameters. Each Bob can be trusted to follow Alice's instructions but they are not guaranteed to be secure from observation. The Bobs do not need to be able to communicate with one other. Alice sends states that are entangled over all of the Bobs or an equal sized set of separable states to all of the Bobs in each round with probabilities $P_E$ and $P_S=1-P_E$ respectively. Each Bob chooses at random to apply a phase to the state they receive with their parameter, sending it along the lower path, or not, sending it along the upper path, with probabilities $P_{M,b}$ and $P_{F,b}=1-P_{M,b}$ respectively. Then they measure them and send the measurement basis, discrete measurement results and whether they applied the parameter to Alice through a public classical communication channel. Alice uses these results to verify for an eavesdropper and estimate the function of parameters.}
    \label{fig:NetworkBasic}
\end{figure}

In this paper we extend the two party protocol in~\cite{Moore2023} to a scenario where a function of parameters distributed across a network of remote Bobs is estimated without using any additional communication channels, as shown in Fig.~\ref{fig:NetworkBasic}. The relative advantage of this protocol is greater for networks of increasing size as the number of extra communications channels that would otherwise be needed grows. We study the limitations of performing SQRS to estimate a function when considering each node separately and when distributing an entangled state across the nodes. We then show how a hybrid protocol using both separable and entangled states can outperform both, achieving quantum-enhanced measurement precision while ensuring strict bounds on information privacy.
 
The paper is structured as follows. In Section~\ref{sec:protocol} we set out the steps to our protocol and use Monte Carlo simulations to demonstrate the effectiveness of using separable or entangled initial states as well as a hybrid protocol. We consider two scenarios: 1. Alice performing a predetermined number of rounds and 2. Eve attacking every round by measuring and replacing the states in the quantum channel with her own entangled states until Alice detects her for the first time and stops the protocol. To maintain security we set a limit on the average information Eve can gain over many simulations and, within this limit, search for the optimal choice of parameters to maximise Alice's estimation effectiveness. We show that entangled initial states are not effective for security as the number of Bobs increases, whereas separable initial states remain secure but have reduced measurement precision. Using a hybrid protocol with some separable and some entangled initial states provides security while performing parameter estimation with some of the quantum enhancement of a global estimation strategy.
 
In Sections~\ref{sec:information_gain} and \ref{sec:security} we demonstrate the information gain and privacy of this hybrid protocol respectively. Section~\ref{sec:information_gain} begins by discussing Fisher information and why the asymptotic large data limit is difficult to reach in a secure network. Section~\ref{sec:security} begins with an analytical calculation of the number of times Eve can successfully intercept information using attacks on the quantum and classical channels before she is detected. The sections then respectively show the information gain for Alice and the limits on Eve's information gain for our hybrid protocol using qubits and generalised GHZ states to measure a function of phase parameters. In particular, these sections show the main result of this paper, namely a practical way to achieve both network security and a measurement precision that surpasses the standard quantum limit.

\section{Protocol}\label{sec:protocol}

Our SQRS protocol proceeds as follows:

\begin{enumerate}
    \item Alice prepares either a separable state to be sent to each Bob or an entangled state to be distributed over all of the Bobs with probabilities $P_S$ and $P_E$ respectively. \begin{enumerate}
        \item If Alice prepares a set of separable states $\ket{S} = \bigotimes_{b=1}^{N_B} \ket{S_b}$, each Bob is sent a Pauli-X or Pauli-Y eigenstate
        \begin{equation} \label{eq:InitialStateS}
            \ket{S_b} = \left( \ket{0} + e^{i\chi_b}\ket{1} \right)/\sqrt{2}
        \end{equation}
        with $\chi_b$ chosen randomly with equal probability from the set $\chi_b \in \{ 0, \pi/2, \pi, 3\pi/2 \}$. The density function from the perspective of an eavesdropper, who does not know what the individual states are, is 
        \begin{align}
            \rho = &\frac{1}{4} \ket{X+}\bra{X+} + \frac{1}{4} \ket{Y+}\bra{Y+} + \nonumber \\  &\frac{1}{4} \ket{X-}\bra{X-} + \frac{1}{4} \ket{Y-}\bra{Y-} = \frac{1}{2}I
        \end{align}
        where I is the identity matrix. Evolution with a unitary operator such as a phase gate does not change this density function indicating that an eavesdropper cannot interpret the measurement results to estimate phase parameters without having more information about the state used in each round.
        \item If Alice prepares an entangled state she sends a particle from a generalised GHZ state to each Bob,
        \begin{equation} \label{eq:InitialStateE}
            \ket{E} = \left( \ket{0}^{\otimes N_B} + e^{i\chi}\ket{1}^{\otimes N_B} \right)/\sqrt{2}
        \end{equation}
        where $\chi \in \{ 0, \pi/2, \pi, 3\pi/2 \}$ at random with equal probability for each state sent. These are generalised GHZ states with Pauli-X and Pauli-Y encoding. Similarly, from an eavesdropper's perspective $\rho = \frac{1}{2}I$ ensuring that an eavesdropper cannot interpret the measurement results to estimate the total phase applied to the probes.
    \end{enumerate}
    \item Alice sends the quantum states through a quantum communication channel to the Bobs with each receiving the state required for their individual parameter estimation. An eavesdropper could perform a spoof attack by applying a phase to the probe or the first step of a man in the middle attack here by measuring and replacing (or simply replacing) the probes in some or all of the rounds.
    \item Each Bob either \begin{enumerate}
        \item with probability $P_M$ applies the (unknown) phase, $\phi_b$, that Alice wishes to measure to his probe using a phase gate $P(\phi_b)$ and measures it in either the Pauli-X or Pauli-Y basis with equal probability. This will aid Alice in estimating that parameter and the function of parameters.
        \item with probability $P_F$ measures his probe  in either the Pauli-X or Pauli-Y basis with equal probability. This may allow Alice to perform a fidelity check and verify for man in the middle attacks.
    \end{enumerate}
    Here, Eve could perform the second step of a man in the middle attack by observing the measurement results of the Bobs.
    \item All the Bobs then communicate their measurement basis, measurement outcome and whether they applied their parameter to the state to Alice through the public classical communication channel. Similar to stage 3 above, Eve could perform the second step of a man in the middle attack by observing the publicly announced results.
    \item Alice uses the measurement results to aid in estimating the function of parameters and verify for man in the middle attacks.\begin{enumerate}
        \item If Alice prepared separable states in step 1a and a Bob applied the phase $\phi_b$ in step 3a and used a measurement in the Pauli-X basis then the probability of the two measurements is
        \begin{align} \label{eq:Pcos}
            P\left(\pm 1 | \chi_b \right) = &\left| \braket{X\pm|P(\phi)|S_b(\chi_b)}  \right|^2 \nonumber  \\ &= \frac{1}{2}\left(1\pm\cos\left(\phi_b+\chi_b\right)\right).
        \end{align}
        If, instead Bob performs measurement in the Pauli-Y basis the probability of the two measurement results is 
        \begin{align} \label{eq:Psin}
            P\left(\pm 1 | \chi_b \right) = &\left| \braket{Y\pm|P(\phi)|S_b(\chi_b)}  \right|^2 \nonumber  \\ &= \frac{1}{2}\left(1\mp\sin\left(\phi_b+\chi_b\right)\right).
        \end{align}
        These measurement results can be used to estimate $\phi_b$ which can then be used to aid in estimating $\theta$.
        \item If Alice prepares separable states in step 1a and Bob measures the state without applying any phase in step 3b Alice can perform a fidelity check on the quantum states to verify for man in the middle attacks on the quantum communication channel. Alice expects measurements in the same orthogonal basis as the initial state to always give a result corresponding to the initial state and measurements in the other orthogonal basis should give each result $1/2$ of the time. As the states are chosen from an indistinguishable set and all of them have some probability of being verified, any man in the middle attack that manipulates the quantum states such as spoofing by applying a phase shift to the quantum states or stealing information by measuring and replacing the quantum states risk changing the state causing a failed fidelity check and allowing Alice to detect the attack.
        \item If Alice prepares entangled states in step 1b and each Bob has chosen to apply their phase or not in step 3a or 3b, the net phase, $\varphi$, is the sum of the phases that the Bobs have applied to the state. Similarly, the net measurement is the sum of the measurement phases applied by the Bobs with result probabilities given by equations~\ref{eq:Pcos} and~\ref{eq:Psin} with $\phi_b\rightarrow \varphi$. There are $2^{N_B}-1$ different $\varphi\neq 0$ which Alice can estimate and use to aid in estimating $\theta$.
        With probability $\prod_{b=1}^{N_B} P_{F,b}$, where $P_{F,b}$ is the probability of the $b^\text{th}$ Bob performing a fidelity check, no Bob applies a phase, $\varphi=0$, allowing Alice to make a single fidelity check using the entire entangled state in the same way as one of the separable states sent to a single Bob in step 5b.
    \end{enumerate}
\end{enumerate}

All SQRS protocols must find a way to balance both security and estimation efficiency. If we assume there is no eavesdropper, any protocol would be most efficient by having fidelity checking probability, $P_F=0$ and parameter measurement probability $P_M=1-P_F=1$. Similarly, the most secure protocol has $P_F=1$ and $P_M=0$. We choose $P_F$ to be equal for all Bobs and likewise for $P_M$, as this optimises the measurement efficiency and the security for the sum of parameters, $\theta = \sum_{b=1}^{N_B}\phi_b$.

For security purposes, the average number of fidelity checks in a round with separable initial states is $\sum_{b=1}^{N_B}P_{F,b}$ which depends only on the sum of the probability of fidelity checks. The probability of a fidelity check in a round with entangled initial states is $\prod_{b=1}^{N_B} P_{F,b}$ which is maximised for fixed $\sum_{b=1}^{N_B}P_{F,b}$ when $P_F$ is the same for all Bobs. Equivalently, from a metrology perspective each parameter contributes equally to providing information for the sum of the parameters, $\theta$, making it logical to have the same $P_M$ for each Bob. As $P_F+P_M=1$ we arrive at the same result that the probability of a fidelity check in a given round with entangled states is maximised when when $P_F$ is the same for all Bobs.

\begin{figure}
    \centering
    \includegraphics[width=.48\textwidth]{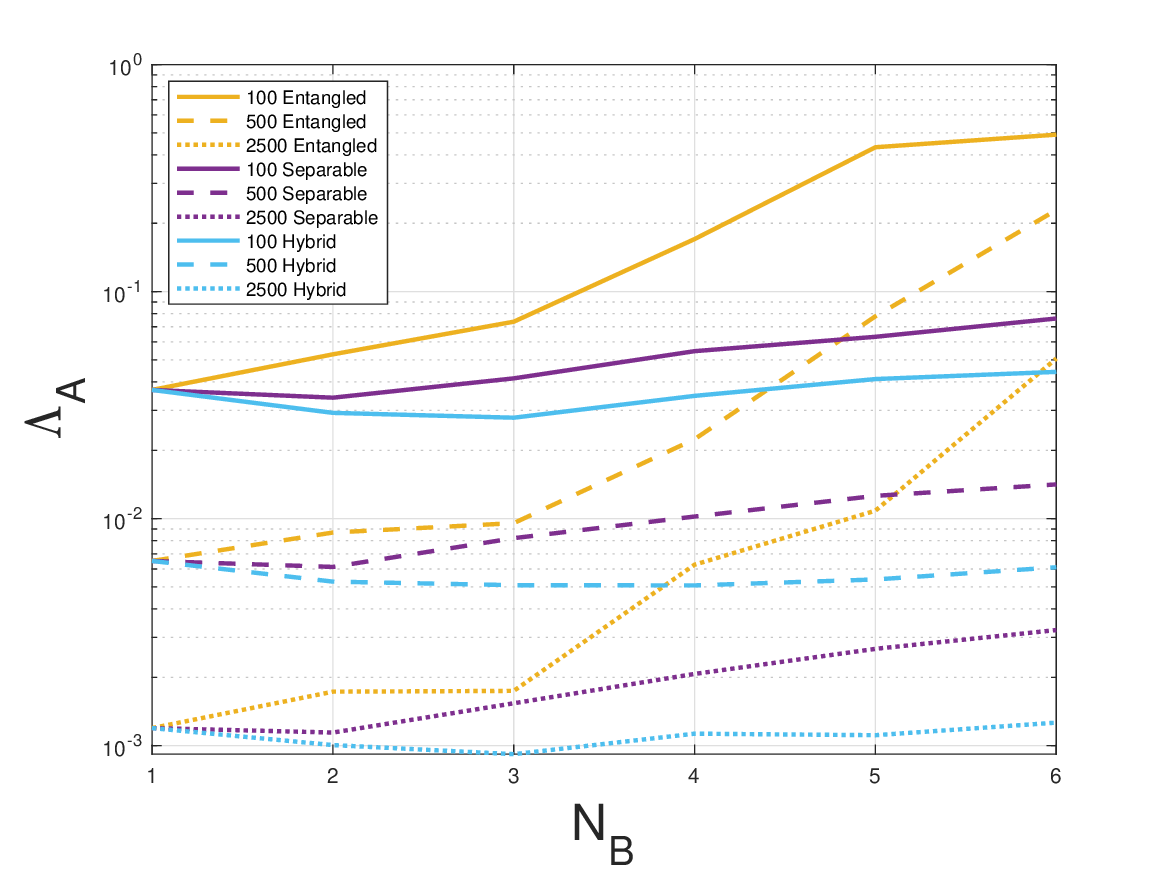}
    \caption{The mean, $\Lambda_A$ of Alice's likelihood function circular mean square error, $\lambda(\vec{n},\vec{\phi})$ averaged over the sets of possible results $\vec{n}$ for many different values of Bobs' phases $\vec{\phi}$ in situations where Eve's mean circular mean square error has a lower bound $\Lambda_E \geq 1/2$ and minimised with respect to the protocol parameters $P_S$ and $P_F$ for a measure and replace with entangled states attack. This is plotted for three approaches: initial states entangled over all Bobs; initial states separable between the Bobs; hybrid of separable and entangled initial states used with probability of separable initial state $P_S$ for each round. Solid lines represent a maximum of 100 rounds, dashed lines for 500 rounds and dotted lines for 2500 rounds.}
    \label{fig:MCAliceDispersion}
\end{figure}

\begin{figure*}
    \centering
    \includegraphics[width=.95\textwidth]{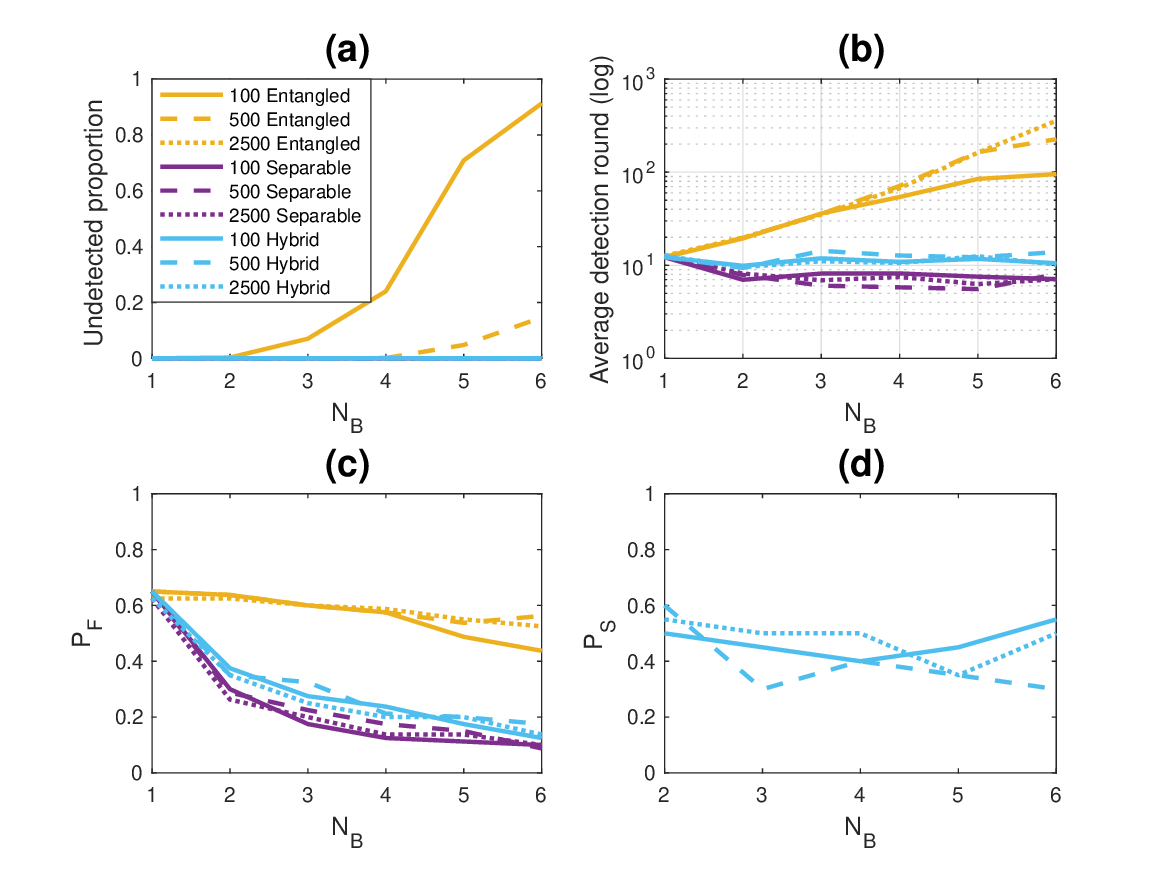}
    \caption{Optimisation results for $\Lambda_E\geq 0.5$ when Eve perform a measure and resend entangled state attack. (a)The proportion of times that Alice does not detect Eve before the end of the protocol. Other than 100 and 1 000 rounds entangled initial states, all of the lines are at or very close to zero and cannot be distinguished on this plot. (b) The number of rounds until Alice detects Eve for the first time. (c) Fidelity checking probabilities $P_F$. (d) Separable state probabilities $P_S$ for the hybrid protocol.}
    \label{fig:Optimisation4ExtrasER}
\end{figure*}

We performed Monte Carlo simulations under two different scenarios to demonstrate the effectiveness of our protocol in the low-data minimal prior information regime for estimating  $\theta$ while maintaining security against an eavesdropper, Eve, who intercepts the states that Alice sends, measures them and replaces them with corresponding entangled states. To properly compare the two different scenarios a consistent measure of information gain is required. In limited data, distributions are non-negligible over a $2\pi$ range so, the circular nature of phase parameters must be accounted for when analysing information gain. A circular measure of information gain equivalent to the mean square error for narrow distributions, where circular effects are negligible, is preferable. We use a measure of dispersion $\lambda(\vec{n},\vec{\phi})$ drawn from the measure of the distance between two angles $C_{\theta,\hat\theta}=1-\cos(\hat\theta-\theta)$ as a function of the true parameter values $\vec{\phi}$ and the number of results $\vec{n}$,

\begin{equation}
    \lambda(\vec{n},\vec{\phi}) = \int_0^{2\pi} \left(1-\cos(\hat{\theta}-\theta) \right) \mathcal{L}(\hat\theta|\vec{n})d\hat{\theta} ,
\end{equation} \label{eq:circularMSE}
Our choice is justified as $C_{\theta,\hat\theta}$ is proportional to the cost function ${2C_{\theta,\hat\theta}=\tilde{C}_{\theta,\hat\theta}=4\sin(\frac{\hat\theta-\theta}{2}) = (\hat{\theta}-\theta)^2} + \mathcal{O}( (\hat{\theta}-\theta)^4)$ used for limited-data phase estimation with minimal prior information, popular due to it being the simplest function that approximates the variance for narrow distributions~\cite{Demkowicz-Dobrzański2011a}. This provides a useful measure where $\lambda(\vec{n},\vec{\phi}) = 0$ corresponds to a delta-function distribution around the true value, $\lambda(\vec{n},\vec{\phi}) = 1$ corresponds to a uniform distribution around $2\pi$, indicating no information in a circular sense and $\lambda(\vec{n},\vec{\phi}) = 2$ corresponds to a delta function around $\theta\pm\pi$, which is the maximum possible error indicating measurement error or spoofing. 

$\lambda(\vec{n},\vec{\phi})$ is a measure of information gain for a single iteration with results $\vec{n}$ for parameters $\vec{\phi}$. To be able to compare the information gain from different scenarios we need to make a fair comparison that is not specific to a single implementation of the protocol. Instead, we use the average of $\lambda(\vec{n},\vec{\phi})$ made over many sets of measurement results $\vec{n}$ from simulations with different sets of true parameters $\vec{\phi}$. We denote this average $\Lambda(N_R,P_S,P_F)$, where $N_R$ is the number of protocol rounds. Our circular data analysis techniques are further discussed in Appendix~\ref{sec:CircularStatistics}.

In the first scenario there is no eavesdropping so we calculate $\Lambda$ for Alice (denoted $\Lambda_A$) as a function of $N_R$. In the second scenario Eve attacks every round measuring and replacing the states with her own states that are entangled over all Bobs and correspond to her measurement results. She continues to do this until Alice detects her the first time or Alice reaches the maximum number of rounds she intended to run the protocol, $N_R$, at which point the protocol is stopped. The security requirements are dependent on the specific scenario. Here we have chosen to limit Eve's value of $\Lambda$, $\Lambda_E$, to $\Lambda_E\geq0.5$, approximately equivalent to a linear mean squared error of at least 1. This gives Alice a guarantee on how little information Eve can gain. The limit put on $\Lambda_E$ can, of course, be varied depending on the scenario and application.

These two scenarios were chosen because the detection of Eve depends on the number of rounds that she attacks, not the number of rounds that Alice attempts. Also, to be sure that Eve could not get as much or more information than Alice, Alice would want to ensure that Eve is detected a long time before reaching the total number of rounds. If Alice reaches the total number of rounds intended and Eve attacks without being detected it should be because she has only attacked a relatively small proportion of the rounds and thus causes only a small perturbation to Alice's results. We attempt to minimise $\Lambda_A$ while ensuring $\Lambda_E\geq0.5$ and that Eve is detected long before the end of the predetermined rounds. $\Lambda_A$ and $\Lambda_E$ are calculated for a range of values of $P_S,P_E\in[0,1]$ with the same amount of averaging for the same choices of $\vec{\phi}$ for each $N_B$. Further details of the methodology can be found in Appendix~\ref{sec:MCMC_optimisation}.

The results of these simulations for 100, 500 and 2500 rounds are shown in Fig.~\ref{fig:MCAliceDispersion}. They demonstrates that a hybrid of separable and entangled initial states outperforms the use of only one of the two, showing little variation in information gain with the number of Bobs. For the remainder of this section we will discuss these results in more detail while comparing them to Fig.~\ref{fig:Optimisation4ExtrasER} which  shows (a) the probability of Eve going undetected, (b) the rounds until detection and (c,d) the protocol probabilities chosen by the optimisation algorithm. More detail on the information gain and security aspects can be found in Sections~\ref{sec:information_gain} and \ref{sec:security} respectively.

Entangled initial states are plotted in yellow (middle grey) on the figures. Fig.~\ref{fig:MCAliceDispersion} demonstrates that this choice of Alice's initial state performs increasingly worse than separable-only initial states and a hybrid protocol as the number of Bobs increases to the point that with low data and larger numbers of Bobs, Alice does little better than the security limits placed on Eve (i.e. $\Lambda\geq0.5$). There are two issues with entangled initial states. The first is that, with limited data, it is difficult to gain much information for certain protocol parameters, further discussed in section~\ref{sec:information_gain}. The second is that security reduces with the size of the network, further discussed in section~\ref{sec:security}.

For low data entangled only initial states performs even worse than Fig.~\ref{fig:MCAliceDispersion} would suggest because the proportion of rounds where Eve goes undetected before the end of the protocols. The results shown in Fig.~\ref{fig:MCAliceDispersion} do not account for Eve going undetected. Fig.~\ref{fig:Optimisation4ExtrasER}(a) shows that 5 or more Bobs with 500 rounds and 3 or more Bobs with 100 rounds there is a non-negligible probability of Eve going undetected for the entire protocol. Firstly, this is unacceptable from a security point of view. Secondly, Fig.~\ref{fig:Optimisation4ExtrasER}(b) demonstrates that this corresponds to an average number of rounds before detection being of the same order of magnitude as the total rounds. This means that the effect of an undetected eavesdropper (who may attack only some rounds) on Alice's estimation would be more than a small amount of noise making her perform even worse than in Fig.~\ref{fig:MCAliceDispersion} or forcing her to perform significantly more fidelity checks than Fig.~\ref{fig:Optimisation4ExtrasER}(c) suggests which would also make her estimation even worse. 

Separable initial states are plotted in dark purple (dark grey) on the figures. These show excellent security features regardless of the number of rounds. It can be seen in Fig.~\ref{fig:Optimisation4ExtrasER}(b) that Alice's optimised information gain is achieved while ensuring that Eve can attack fewer than 10 rounds on average before she is detected, independent of the number of rounds or number of Bobs. For separable states, the number of fidelity checks for a constant $P_F$ increases linearly with the number of Bobs so, as demonstrated in Fig.~\ref{fig:Optimisation4ExtrasER}(c), $P_F$ can be reduced as the number of Bobs increases. However, the disadvantage of using separable initial states is that we miss out on the quantum enhanced measurement precision of entangled states. We see in Fig.~\ref{fig:MCAliceDispersion} that this causes the measurement efficiency to reduce with the number of Bobs. The reduction is close to the linear reduction in a metrology protocol without security.

Hybrid initial states are plotted in light blue (light grey) on the figures. As the number of Bobs increases, the security is increasingly reliant on the separable initial states, Fig.~\ref{fig:Optimisation4ExtrasER}(d). This is because the average number of security checks per round for entangled initial states is given by $P_F^{N_B}$ and so reduces rapidly with network size. By contrast, the average number of security checks per round for separable initial states increases with network size  as $N_BP_F$ since there is the possibility of more than one check per round. The fidelity checking probability for hybrid initial states when optimised to fulfil security conditions and minimise Alice's estimation dispersion is shown in Fig.~\ref{fig:Optimisation4ExtrasER}(c) and follows a similar shape to the separable initial states. Similar to the separable only states, the average number of rounds before Eve is detected remains fairly low for any number of Bobs. It is not quite as low as the case of separable only states because, as shown in Fig.~\ref{fig:Optimisation4ExtrasER}(d), many rounds make use of entangled states which have reduced security as discussed above. However, it remains sufficiently small that, in cases where Eve does not make enough attacks to be detected, we can approximate Alice's information as that given by a protocol with no eavesdropper.

\begin{figure}
    \centering
    \includegraphics[width=.48\textwidth]{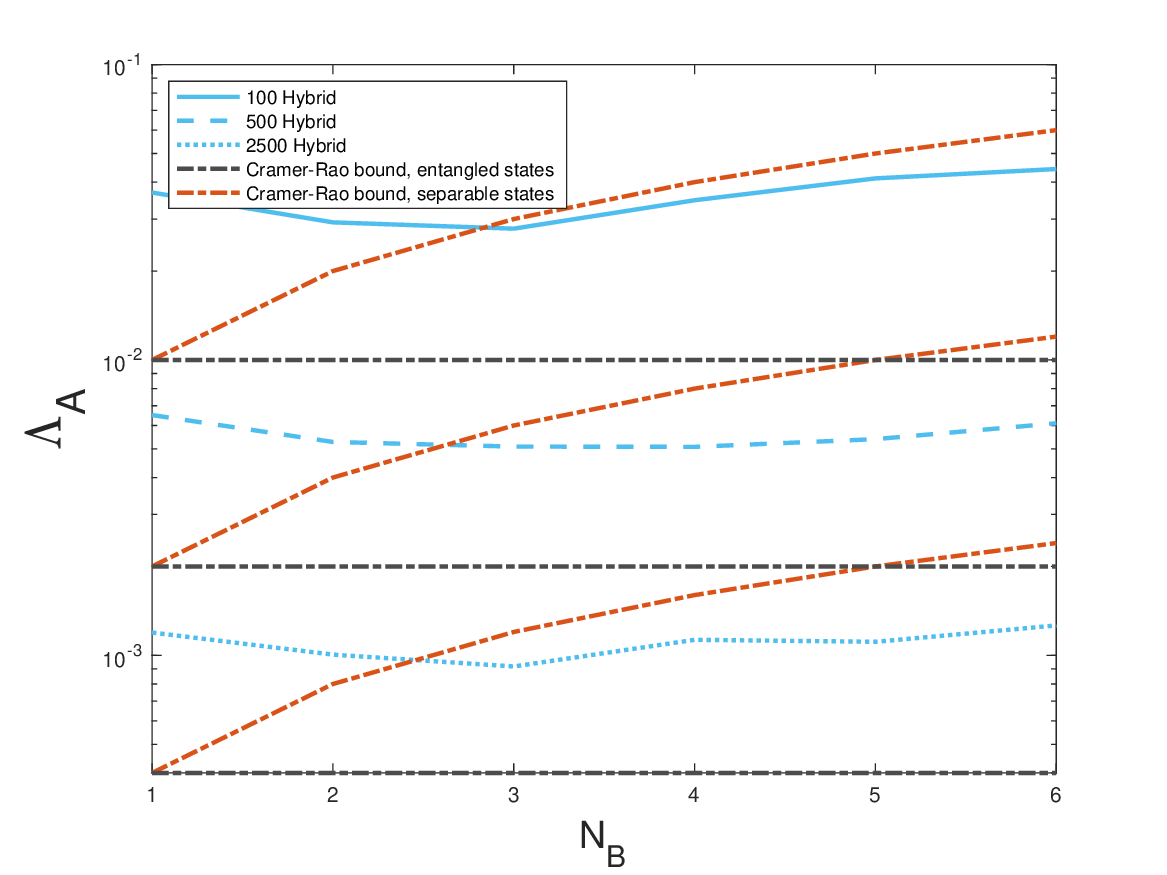}
    \caption{Comparison of Alice's maximum $\Lambda_A$ while limiting Eve's  $\Lambda_E\geq0.5$ with the Cramér-Rao bound for non-secured metrology protocols using separable and entangled probes.}
    \label{fig:MCDispersionCRB}
\end{figure}

The Cramér-Rao bounds for similar metrology protocols without security for separable and entangled initial states have variances given by $N_B/N_R$ and $1/N_R$ respectively, where $N_R$ is the number of rounds. When using hybrid initial states there is a trade-off between the enhanced security of separable states and enhanced measurement precision of entangled states. Fig.~\ref{fig:MCDispersionCRB} shows that for three or more Bobs, the hybrid protocol (with $\Lambda_E\geq 0.5$) has an average dispersion less than the Cramér-Rao bound for separable probes and 3 or more Bobs. This shows quantum enhanced measurements and security combined into a single protocol.

\section{Information gain in the asymptotic and low-data limits} \label{sec:information_gain}

The information Alice and Eve gain can be considered in the same way and depends on the number of rounds $N_R$ and the protocol parameters $N_B$, $P_S$ $(P_E=1-P_S)$ and $P_F$ $(P_M=1-P_F)$. The results in this section will be given in terms of these four independent parameters: $N_R$, $N_B$, $P_S$ and $P_F$. If Eve performs an attack where she replaces states with her own entangled or separable states then the results correspond to $P_S=0$ or $P_S=1$ respectively. 

In a protocol with $N_B$ Bobs each measuring a parameter $\phi_b$, the set of combinations of Bobs that could perform parameter measurement or not in each round is of size $2^{N_B}$; there are $\binom{N_B}{m}$ combinations for $m$ parameters being measured. We write the set of possible combinations $\vec{\varphi} = \{0,\phi_1,\phi_2,...,\phi_1+\phi_2,\phi_1+\phi_3,...,\theta\}$ with each $\varphi_k(m)$ an element of $\vec{\varphi}$ with $m$ parameters measured. Appendix~\ref{sec:FisherInformation} demonstrates how the Fisher information of the protocol can be calculated by combining the information due to the sets of $\varphi_k(m)$ for each $m$. The probability of each $\varphi_k(m)$ being measured is 

\begin{equation} \label{eq:probK}
    P_k(m)  = P_SP_M\Big|_{m=1} + P_EP_M^mP_F^{N_B-m}.
\end{equation}
Assuming that the Fisher information of each $\varphi_k(m)$ is the same for any given value of $m$, the Fisher information of the entire protocol relative to round total round count is

\begin{equation}
    \mathcal{I}_\text{total} = P_S\frac{ P_M\mathcal{I}_1 }{N_B} + P_E\sum_{m=1}^{N_B} P_M^m P_F^{N_B-m} \frac{m}{N_B}\binom{N_B-1}{m-1}  \mathcal{I}_m,
\end{equation}
where $\mathcal{I}_m$ is the classical Fisher information of each $\varphi_k(m)$. For clarity, the entangled and separable state contributions have been separated. The entangled states contribute a term for $m=1,2,...N_B$ and the separable states contribute a term for $m=1$. The same relationship also holds for the corresponding quantum Fisher information.

The Fisher information is related to the estimation uncertainty with the Cramér-Rao bound,

\begin{equation} \label{eq:CRBtheta}
    (\delta \theta)^2 \geq \frac{1}{N_R \mathcal{I}_{total}}.
\end{equation}
This provides an lower bound to the estimation of $\theta$ in the asymptotic limit of large $N_R$. In the low data limit, when prior information is accounted for, this bound is not valid and can be surpassed~\cite{Rubio_2018,Rubio_2019,Rubio_2020a,Sidhu2020a,meyer2023quantummetrologyfinitesampleregime}. The prior information used in this article is discussed at the end of this section.

Suppose that $N_\mathrm{CR}$ is the number of phase measurements (of the form given in equations~\ref{eq:Pcos} and~\ref{eq:Psin}) to be assured we are in the regime where the Cramér-Rao bound is valid for any $\varphi_k(m)$. If there are $N_\mathrm{CR}$ measurements for each $\varphi_k(m)$ then, given  equation~\ref{eq:CRBtheta}, the Cramér-Rao bound for $\theta$ should also be valid. However, there are $2^{N_B}$ different $\varphi_k(m)$ which grows rapidly with $N_B$. Apart from the special cases $P_E=0$ or $P_F=0$ these will all have a non-zero probability of occurring. If $P_E=1$ and $P_F=0.5$ they are all equally likely to occur, doing so with probability $2^{-N_B}$; for any other protocol parameters some will be even less likely. Therefore at least $2^{N_B}N_\mathrm{CR}$ rounds are required to assure the validity of equation~\ref{eq:CRBtheta}, which grows rapidly with $N_B$. This signifies that it is important to pay close attention to the amount of information gain in limited data using a measure such as $\Lambda$ rather than being overly reliant on the Fisher information, especially in networks with multiple Bobs.

This can be mitigated if we consider that for some choices of system parameters the $\varphi_k(m)$ for some values of $m$ are very unlikely to occur so, they would contribute very little to the information gain. However, it is not advantageous to consider only the system parameters that reduce the number of $\varphi_k$ that need to be considered. For instance, taking $P_F=0.1$ or $P_F=0.9$ many of the $m$ are quite unlikely but they have low Fisher information and security respectively; while $0.1<P_F<0.9$ increases the number of $\varphi_k$ that need to be considered, if there are enough rounds it provides better information gain than $P_F=0.1$ and it always provides better security than $P_F=0.9$, so as demonstrated in Fig.~\ref{fig:Optimisation4ExtrasER}(c) less extreme values of $P_F$ provide good balance between the two effects.

When the data are limited, and $\binom{N_B}{m}$ is large, it is not effective to perform parameter estimation of $\theta$ by combining all of the $\varphi_k(m)$ for each $m$ together. We can illustrate this by considering how the variance of the estimator of a random variable is inversely proportional to the number of measurements and the variance of an estimator of a sum of variables is equal to the sum of their variances. This means that the variance of the sum of parameters is limited by the parameter for which we have the least data. For example, if there are no data for one parameter and it has a $2\pi$ range uniform prior, as phases are circular, we gain no information about $\theta$ from all of the other results.

The number of results for all of the $\varphi_k$ is distributed as a multinomial and the marginal of each $\varphi_k$ is a binomial. When there are a lot of $\varphi_k(m)$ compared to the number of rounds the mean number of results for each is small and the standard deviation is large. This makes it inefficient to combine all $\binom{N_B}{m}$ to estimate $\binom{N_B-1}{m-1}\theta$ for limited data. Instead, we gain more information by combining those that have the most results and sum to some multiple of $\theta$. Our limited data parameter estimation methodology is further discussed in Appendix~\ref{sec:InformationGainOptimisation}.

In Fig.~\ref{fig:informationGainQuantumAdvantage} we plot $\Lambda_A$ for 1-6 Bobs between 1 and 2500 rounds for protocols with no fidelity checking and either separable or entangled initial states only compared to the hybrid protocol as optimised for 2500 rounds. It clearly demonstrates that the hybrid protocol, while being secure, is also capable of performing quantum enhanced measurements for functions of parameters spread across a network of sensors with increasing effectiveness with network size. Using the values for $P_F$ and $P_S$ for the hybrid protocol optimised for 2500 rounds ensures that the protocols conform to the security limit up to 2500 rounds. However, it is not optimised for information gain with fewer rounds. This could be further enhanced by combinations of single and multiple pass estimations of $\theta$ set out for two-party SQRS~\cite{Moore2023}.

\begin{figure*}
    \centering
    \includegraphics[width=.95\textwidth]{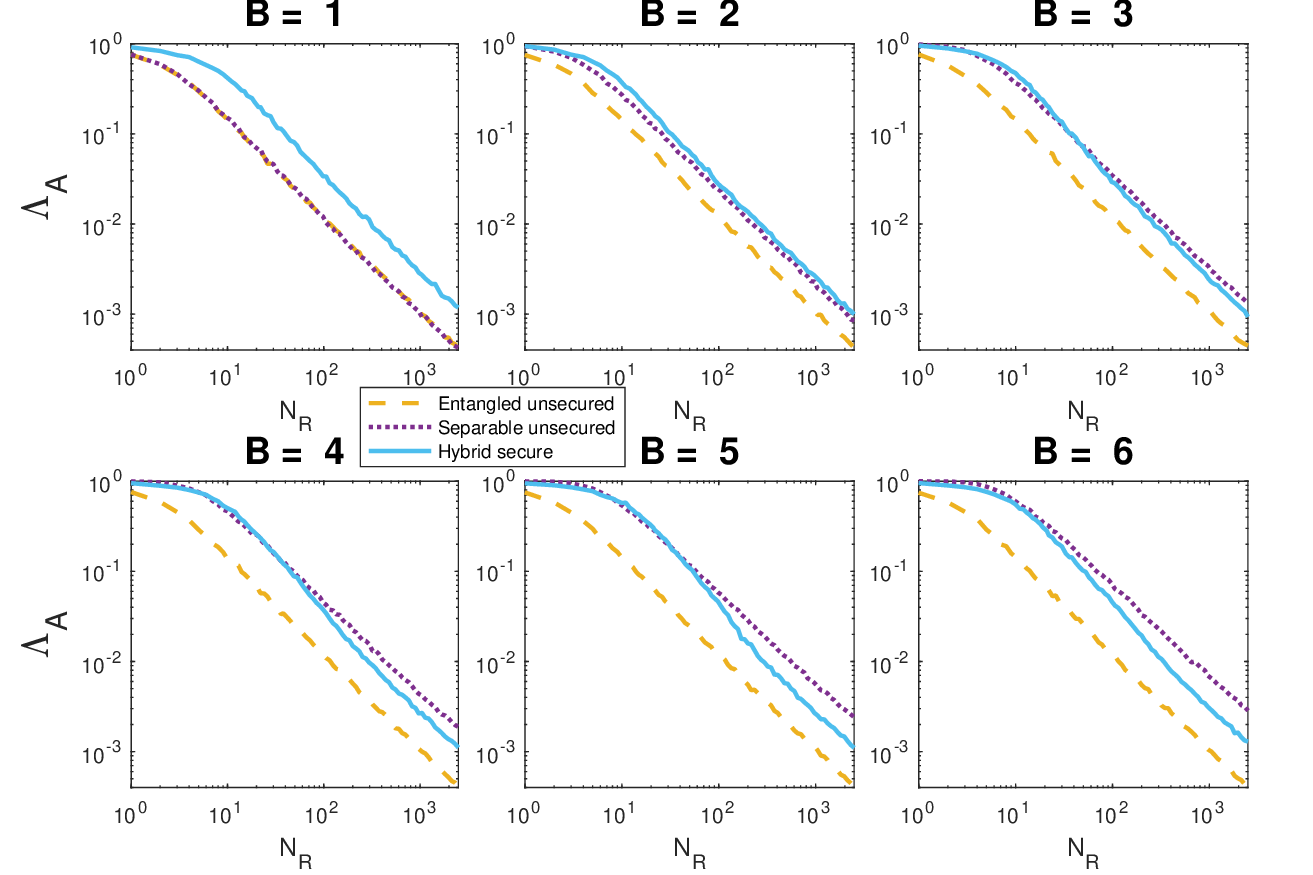}
    \caption{$\Lambda_A$ as a function of the number of rounds, $N_R$, for different protocols and different numbers of Bobs. The purple (dark grey) dotted line is the result for an ideal system using separable states without security. This represents the standard quantum limit using a local estimation strategy which scales as $N_B/N_R$. The yellow (middle grey) dashed line shows the result for an ideal system using entangled states without security. This scales as $1/N_R$. The solid blue line (light grey) is the result for a system that uses a combination of separable and entangled initial states and is secure up to $N_R=2500$. It shows better estimation precision than a local estimation strategy while providing security. The protocol parameters have been taken from Fig.~\ref{fig:Optimisation4ExtrasER} for $N_R=2500$. The information gain has not been optimised for fewer rounds but security is guaranteed.}
    \label{fig:informationGainQuantumAdvantage}
\end{figure*}

The results given in this article use the simplest form of minimal prior information for the estimation of $\theta$, a uniform prior on an arbitrary continuous $2\pi$ range, $p(\theta|\alpha) = \frac{1}{2\pi}$. This prior information provides a posterior distribution equal to the likelihood function over the same range. They can be used interchangeably in equation~\ref{eq:circularMSE} to give the same results. Due to the circular nature of phase parameters, the use of a single prior for $\theta$ leads to indistinguishable parameter estimation in a continuous range no greater than $2\pi$. However, prior information for the individual $\phi_b$ (or a set of $\varphi(m)$ for some $m$) can be used to extend the dynamic range on which $\theta$ can be estimated. No matter the distribution of the priors, separable initial states can be used for the indistinguishable estimation, at best, in an arbitrary $2\pi$ range for each $\phi_b$. Therefore, their sum can be used to estimate $\theta$ in a $2\pi N_B$ range. Keeping track of the estimates of the individual phase parameters, $\phi_b$, becomes particularly useful when considering situations where they may be dynamic. For instance, when estimating only $\theta$ and not analysing the $\phi_b$ data separately a shift of $2\pi/N_B$ in the value of all of the $\phi_b$ would change the value of $\theta$ by $2\pi$ without indicating the change. If, instead, each $\phi_b$ is also estimated and analysed, then such changes would be noticeable. This highlights a further advantage of the hybrid protocol over using only entangled states.

\section{Information asymmetry between Alice and Eve} \label{sec:security}

\begin{figure}
    \centering
    \includegraphics[width=.48\textwidth]{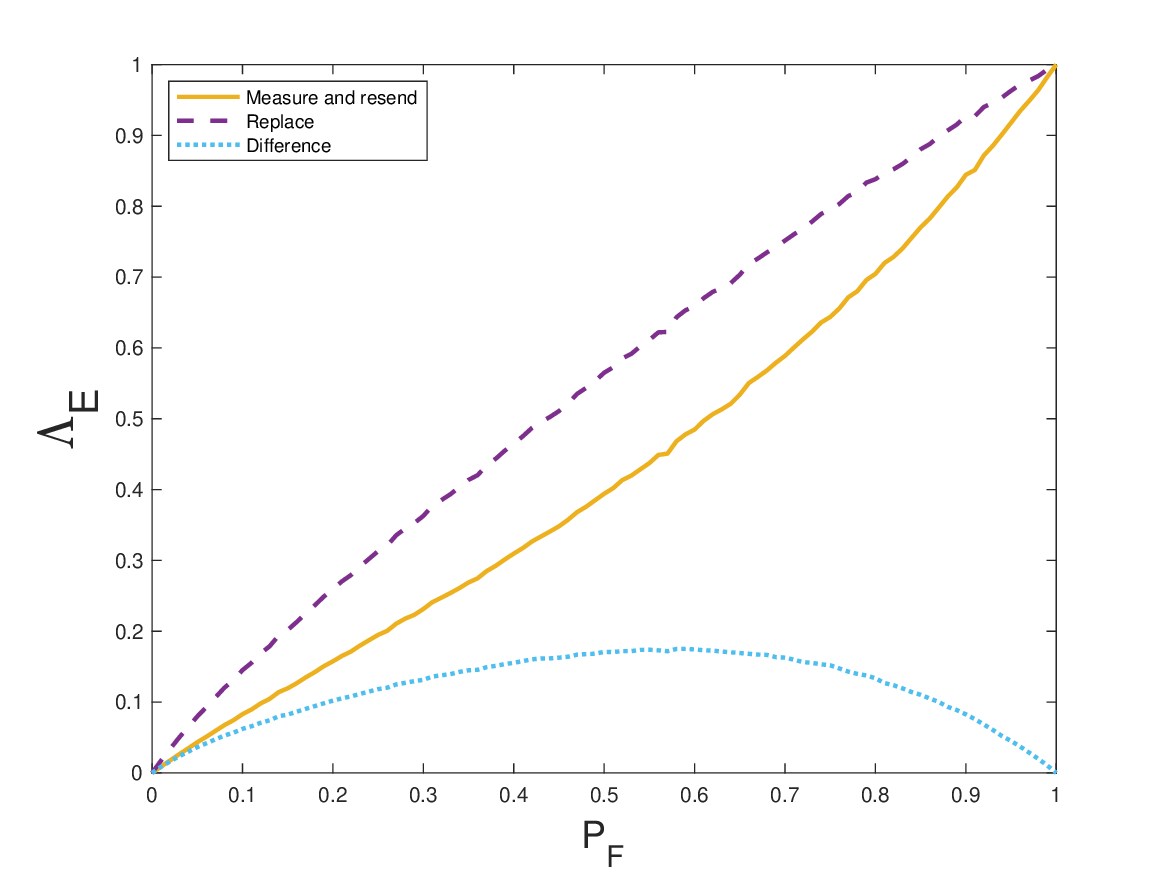}
    \caption{$\Lambda_E$ for a single-Bob protocol with measure and resend attacks and replace attacks. Data was created using simulations of $\Lambda(N_R)$ for $N_R=\{0,1,2,...100\}$ and combining them with the distribution of $N_R$ given by equation~\ref{eq:G1} with $d=F/8$ and $d=F/4$ respectively. For most values of $P_F$ the probability of being undetected in more than $100$ rounds is negligible meaning the values shown in this plot are the true values of $\Lambda_E$. For $P_F<0.1$ the probability is not negligible so the values shown in the plot represent a lower bound on $\Lambda_E$.}
    \label{fig:Security1B}
\end{figure}

Since our protocol has discrete detection results where Eve is either detected or not, we can model the number of rounds until Eve is detected using the geometric distribution. The probability that there are $N_R$ rounds before Eve is detected is given by,

\begin{equation}
    P_\mathrm{Geo1} = (1-d)^{N_R} d \label{eq:G1}
\end{equation}
where $d$ is the probability that Eve will be detected in any given round. This can be used to model the amount of information gain for scenarios with a single Bob. Fig.~\ref{fig:Security1B} shows a lower limit on the amount of information gain for measure and resend attacks and replace attacks on a single Bob. This is also a demonstration of the security in a scenario where Eve attacks to gain information about the phase held by only one of the Bobs. This is calculated by using many Monte Carlo simulation to find $\Lambda$ for 0 to 100 rounds of the protocol then weighting this using the geometric distribution for the number of rounds before Eve is detected. When $P_F$ is very small, there is a non-negligible probability that more than 100 rounds can pass before Eve is detected. As stated in section~\ref{sec:protocol}, this already shows a failure of security. In these cases we set $\Lambda(N_R)=0$ making the plot a lower limit of $\Lambda_E$ for eavesdropping. 

For a single Bob, the detection probability when Alice and Bob verify a qubit on which Eve has performed a measure and resend attack is $1/4$~\cite{Huang2019a,Moore2023}. By a similar logic, when performing a replace attack the state on arrival is random giving a $1/2$ probability of Eve being detected. There are two measurement bases for the initial states and fidelity checks which all occur with equal probability so, there is a $1/2$ probability that Bob performs a fidelity check in the same basis as the initial state. Therefore, the detection probability for a single attacked round is $d(\text{measure and resend}|N_B=1) =p_F/8$ for measure and resend attacks and $d(\text{replace}|p_B=1) = p_F/4$ for replace attacks. This can be generalised to allowing more than one detection using the negative binomial distribution. 

For multiple Bobs we count the number of rounds including the one where Eve is detected. The probability of this is given by another form of the geometric distribution,

\begin{equation}
    P_\mathrm{Geo2} = (1-d)^{N_R-1}d. \label{eq:G2} 
\end{equation}

If Alice only uses GHZ states, then the number of rounds that Eve gains information from, $N_R$, is distributed by equation~\ref{eq:G1}. If Alice uses a separable state, the number of rounds that she gains information on can be described by either equation~\ref{eq:G1} or equation~\ref{eq:G2} depending on whether she gained any information from the final round. A separable state will have several independent tests and measurements and making it possible for Eve to gain some information as well as being detected one or more times in a single round. The maximum information gained for separable states is therefore bounded by what she would get from the number of rounds distributed by equations~\ref{eq:G1} and \ref{eq:G2} from below and above respectively. 

In a protocol using both separable and entangled initial states, we can model the upper limit on the number of rounds from which Eve gains information before she is detected $K$ times by combining the two geometric distributions into a special negative trinomial as a distribution limiting the number information gaining measurements before detection from above,

\begin{equation} \label{eq:SNT}
    P_{SNT}(K) \leq \sum_{k=0}^{\min(N_R-1,K)} \binom{K}{k}u^{N_R-k}\, d_S^k\, d_E^{K-k},
\end{equation}

where $u= 1-d_S-d_E$ is the probability of being undetected in each round, $d_S$ and $d_E$ are respectively the probabilities of being detected from a separable or an entangled state in each round and $N_R$ is the number of measurements that Eve could have got results for before being detected. Similar to the single Bob case, this can be used with values of $\Lambda(N_R)$ to put a lower bound on $\Lambda_E$, an upper bound on Eve's average information gain given the protocol parameters.

The detection probability depends both on the type of attack that Eve performs and the initial state. We take the probability that Eve is detected on a single measurement, when her strategy is to replace Alice's state with a random one, as $d_r$. With initial entangled states, we need a coincidence of all the Bobs performing a fidelity check, $p=P_F^{N_B}$ and the net measurement to be in the same basis as the original state $p=1/2$ in order to detect Eve. The probability of a replace attack being detected when the initial state is entangled, $d(r|E)$, is given by

\begin{equation}
    d(r|E) = d_rP_F^{N_B}/2.
\end{equation}

It is possible to detect Eve more than once in a single round when separable initial states are used. We are interested in the probability of there being at least one detection. For a replace attack the probability a detection by a single Bob is dependent on $d_r$, $P_F$ and the probability that the Bob performs the fidelity check in the same basis as the initials state $p=1/2$. So the probability of a replace attack being detected at least once in a round when the initial state is separable, $d(R|S)$, is given by

\begin{equation}
    d(r|S) = \left(1-(1-d_rP_F/2)^{N_B}\right).
\end{equation}

If Eve measures the quantum states and replaces them with her best guess at the same kind of state, the probability of her being detected, $d_{mr}$, in a fidelity check is less than it would have been if the replacement state was not informed by the measurement outcome, i.e. $d_{mr}\leq d_r$. So, the probabilities of being detected on a single round for measure and replace attacks using separable states to replace separable states, $d(mrS|S)$ and an entangled state to replace and entangled state, $d(mrE|E)$, are given by

\begin{equation}
    d(mrS|S) = \left(1-(1-d_{mr}P_F/2)^{N_B}\right)
\end{equation}

and 

\begin{equation}
    d(mrE|E) = d_{mr}P_F^{N_B}/2
\end{equation}

respectively. If Eve measures an entangled state and replaces it with a separable state, when all of the Bobs perform a fidelity check Alice will interpret the state as if it was an entangled state. The net phase of the set of separable states sent by Eve would be the same as the phase if she had measured and replaced with an entangled state so the detection probability on a single round where Eve measures an entangled state and replaces it with a separable state, $d(mrS|E)$, is the same as if she replaced it with an entangled state,

\begin{equation}
    d(mrS|E) = d_{mr}P_F^{N_B}/2. 
\end{equation}

If Eve measures a separable state and replaces it with an entangled state each Bob has an equal probability of getting a measurement result that corresponds to each of the initial states. The results are dependent on each other and the state that Eve sent but their distribution is the same for each Bob no matter the state Eve sent and no matter the initial state that Alice sent. Therefore, the detection probability when measuring a separable state and replacing it with an entangled state, $d(mrE|S)$, is the same as a replace attack on a separable state,

\begin{equation}
    d(mrE|S) = \left(1-(1-d_rP_F/2)^{N_B}\right).
\end{equation}

\begin{table*}[]
    \centering
    \begin{tabular}{ |p{5cm}||p{4cm}|p{4cm}|  }
 \hline
 \multicolumn{3}{|c|}{Detection probability in each round} \\
 \hline
 Eve attack type & separable initial state & entangled initial state  \\
 \hline
 resend random separable   & $P_S\left(1-(1-d_rP_F/2)^{N_B}\right)$  & $P_Ed_rP_F^{N_B}/2$  \\
 resend random entangled&   $P_S\left(1-(1-d_rP_F/2)^{N_B}\right)$  & $P_Ed_rP_F^{N_B}/2$    \\
 measure and resend separable & $P_S\left(1-(1-d_{mr}P_F/2)^{N_B}\right)$ & $P_Ed_{mr}P_F^{N_B}/2$  \\
 measure and resend entangled & $P_S\left(1-(1-d_rP_F/2)^{N_B}\right)$ & $P_Ed_{mr}P_F^{N_B}/2$  \\
 \hline
\end{tabular}
    \caption{The probability of Eve being detected at least once in a round.}
    \label{tab:detectionProbabilities}
\end{table*}

\begin{figure}
    \centering
    \includegraphics[width=.48\textwidth]{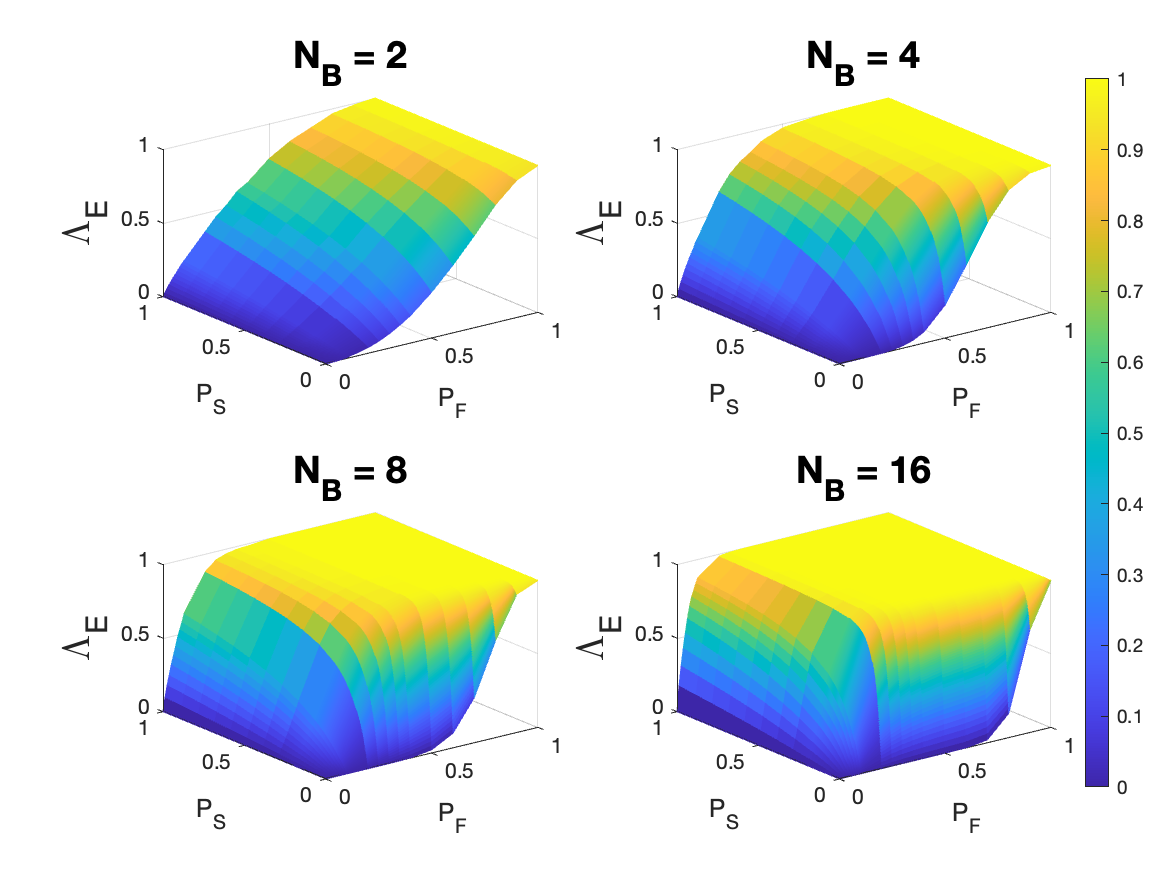}
    \caption{A lower limit on $\Lambda_E$ for 2, 4, 8 and 16 Bobs when performing a measure and replace with entangled state attack. The greater the value, the more secure the protocol.}
    \label{fig:SecurityE4differentBs}
\end{figure}

\begin{figure}
    \centering
    \includegraphics[width=.48\textwidth]{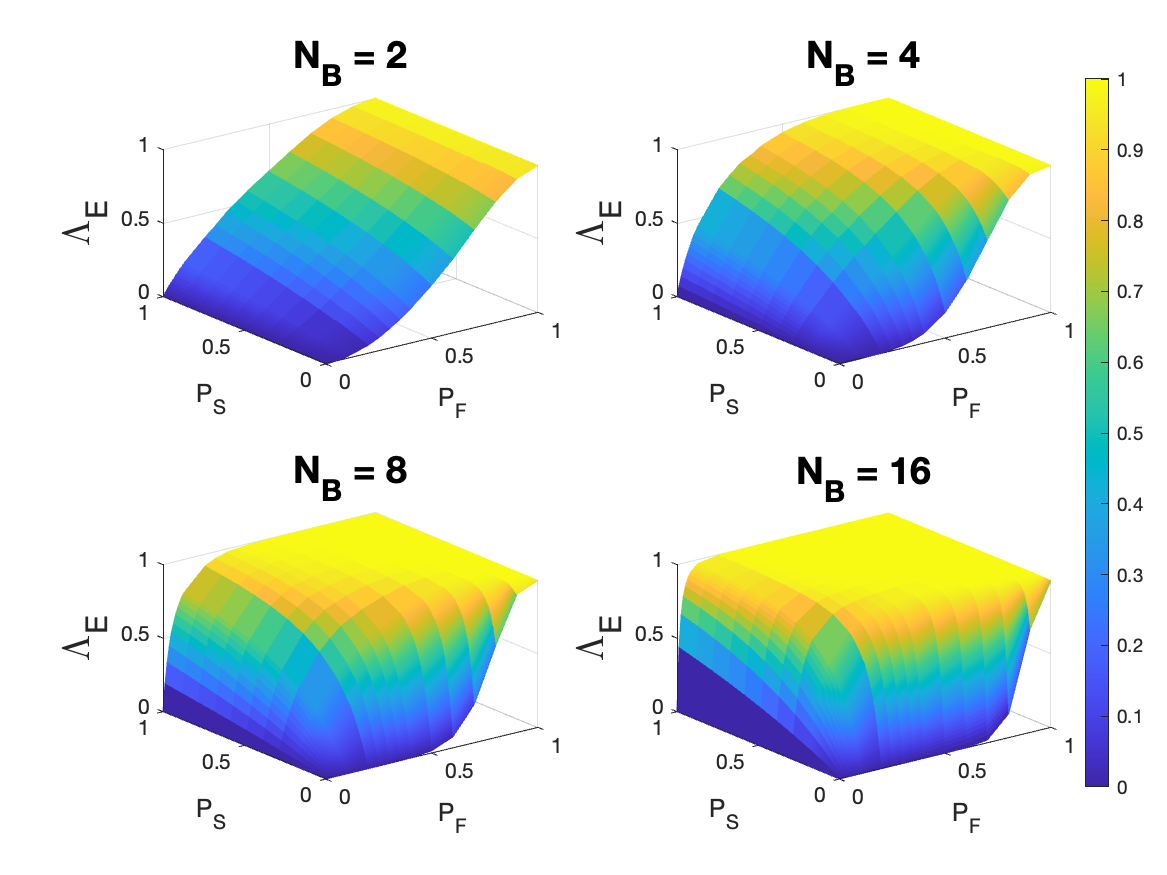}
    \caption{A lower limit on $\Lambda_E$ for 2, 4, 8 and 16 Bobs when performing a measure and replace with separable states attack. The greater the value, the more secure the protocol.}
    \label{fig:SecurityS4differentBs}
\end{figure}

These probabilities are summarised in Table~\ref{tab:detectionProbabilities}. As $d_{mr}\leq d_r$, it is clear that it is always advantageous for Eve to use a measure and resend attack rather than a replace attack. The detection probability is lower when replacing with a separable state than an entangled state but, the information gain is less. Therefore, it is not immediately obvious which attack is better for Eve. However, as the number of Bobs increases, a secure protocol increasingly relies on separable states for security, making the advantage of using measure and replace entangled over replace entangled reduce rapidly with the size of a secure network.

In our protocol Alice stops the first time she detects Eve. Substituting $K=1$ into equation~\ref{eq:SNT}, we get
\begin{equation}
    P_{SNT}(1) \leq u^{N_R-1}\Big|_{N_R\geq1}\, d_S + u^{N_R}\, d_E,
\end{equation}\label{eq:InformationGainingMeasurementLimit}
as a distribution that limits the number of rounds where Eve gains information before she is detected from above. Like the $B=1$ case, $d_r = 1/2$ and $d_{mr} = 1/4$. These, combined with the equations in Table~\ref{tab:detectionProbabilities} allow us to determine this distribution in terms of the system parameters. Similar to the single Bob scenario in Fig.~\ref{fig:Security1B}, using Monte Carlo simulation to find $\Lambda(N_R)$ for $N_R\in\{0,1,2,...50\}$ and using a limiting value $\Lambda(N_R>50)\geq0$ and weighting by the probability distribution of equation~\ref{eq:InformationGainingMeasurementLimit} to put a lower limit on $\Lambda_E$ for Eve's attacks. 

Figs.~\ref{fig:SecurityE4differentBs} and~\ref{fig:SecurityS4differentBs} show this lower limit for 2, 4, 8 and 16 Bobs for measure and resend entangled and separable states respectively. When deciding the parameters $P_S$ and $P_F$ for an implementation of the protocol Alice may choose a security limit $\Lambda_E$ then use any values on these plots that is greater than $\Lambda_E$ to get that level of security. If the intended number of rounds is very large the (inverse) Fisher information can be used to choose the optimal value. However, as discussed in section~\ref{sec:information_gain}, the intricacies of limited data information gain in these scenarios mean that the Fisher information may not be appropriate and it may be better to use some other method to choose optimal values. Either results of optimisation algorithms such as those in section~\ref{sec:protocol} can be used or, as the secure region is already determined, it suffices to perform a Monte Carlo simulation for the number of rounds Alice wants to use to determine a good choice of $P_S$ and $P_F$ selected from those values near the security limit calculated here.

\begin{figure}
    \centering
    \includegraphics[width=.48\textwidth]{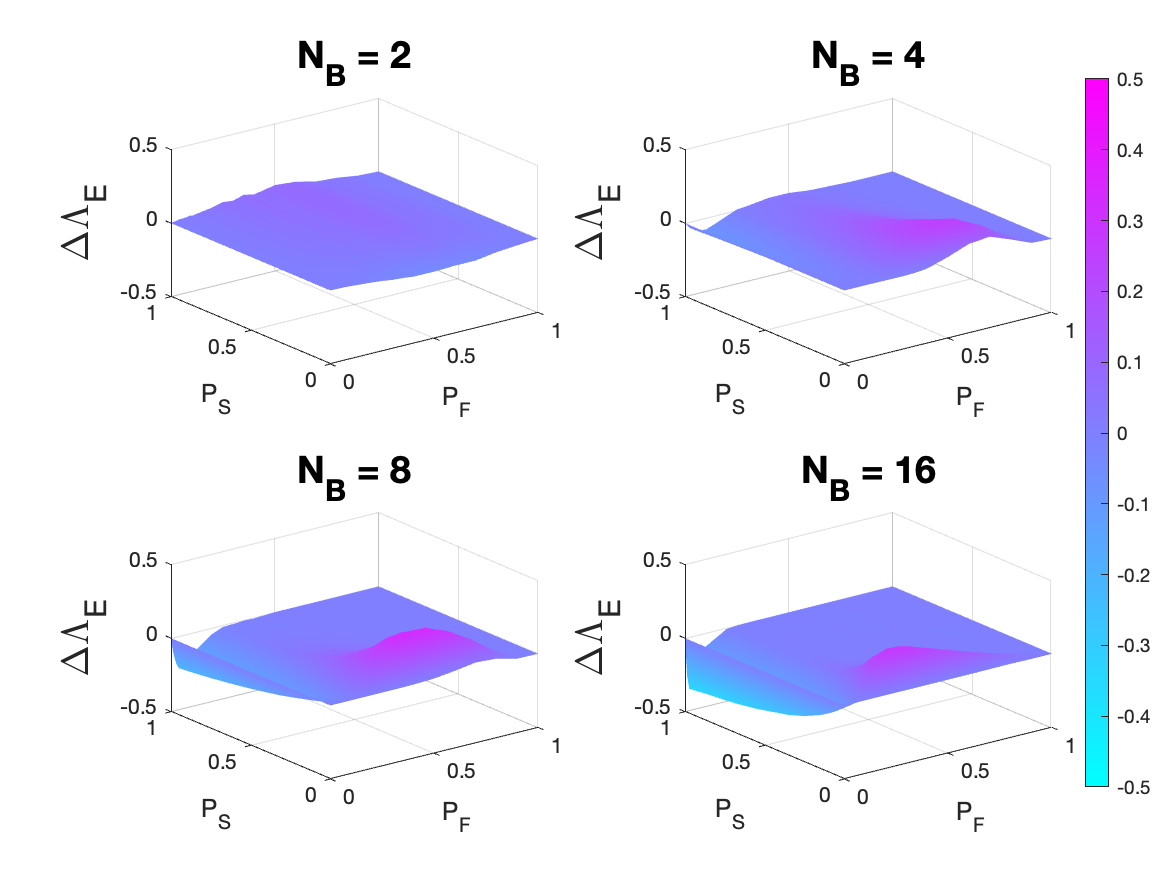}
    \caption{The difference in security between the two attacks in Figs.~\ref{fig:SecurityE4differentBs} and~\ref{fig:SecurityS4differentBs}, $\Lambda_E$(entangled)$-\Lambda_E$(separable). The difference is negligible for the high security choices of S and F indicating that the choice of attack type makes little difference. With lower security $P_S$ and $P_F$ entangled attacks show an advantage for low fidelity while separable states show an advantage for low separable state probability. These differences increase with the number of Bobs.}
    \label{fig:enter-label}
\end{figure}

\section{Conclusion}

We have demonstrated a method of performing quantum-enhanced metrology for functions of phase parameters at a collection of remote sites which is also secure from eavesdropping. The security persists even when the eavesdropper has access to the measurement results, the information in all classical communication channels and the ability to measure and manipulate states in quantum communication channels with the aim of stealing or spoofing information. Furthermore, we have demonstrated that with three or more Bobs the protocol can perform parameter estimation with limited data beyond the standard quantum limit achieved with local estimation strategies while ensuring a fixed privacy limit of $\Lambda_E\geq0.5$.

We have demonstrated how the data can be analysed and the information gain quantified for both Alice, $\Lambda_A$, and Eve, $\Lambda_E$, with limited data. This is done by means of an analytical probability distribution for modelling the number of information gaining measurements Eve can get before being detected. This  is used to put an upper limit on Eve's information gain (lower limit on $\Lambda_E$) before she is detected and enables us to choose protocol parameters $P_S$ and $P_F$ to maximise Alice's information gain (minimise $\Lambda_A$) for any desired privacy limit. 

Our results show a way of implementing quantum enhanced sensing of functions of parameter over remote networks with information privacy and how to optimise this for the linear sum of parameters $\theta = \sum_{b=1}^{N_B}\phi_b$. The performance could be further enhanced by using multipass protocols, quantum-encoded authentication and extended to security against joint attacks on both the classical and quantum channels~\cite{Moore2023}. The protocol may also be adapted to further metrology scenarios such as different prior distributions~\cite{Rubio_2018,Rubio_2019,Rubio_2020a,Sidhu2020a}, noisy scenarios~\cite{Yin2020a,Okane2020a} and be optimised for more complex linear functions~\cite{proctor2017networked,Proctor2018} and non-linear functions of parameters~\cite{Qian2019a,Valeri2020}. 

\section{Data Availability}

Matlab code for the production of data and its analysis is available at the github repository S-W-Moore.

\section{Acknowledgements}

We acknowledge financial support for this research from DSTL under Contract No. DSTLX1000146546.

\section{Author declarations}

The authors have no conflicts to disclose.

\appendix 

\section{Data analysis} \label{sec:CircularStatistics}\label{sec:InformationGainOptimisation}

Linear statistics use a linear support whereas directional statistics, in general, are performed by considering each data point or point on a probability distribution as a vector in a higher dimensional space~\cite{DirectionalStats}. On a circle each data point can be represented by a vector, $\boldsymbol{x}$, parameterised by an angle, $\theta$
\begin{equation}
    \boldsymbol{x} = (\cos(\theta),\sin(\theta))^T,
\end{equation}
or equivalently by a complex number 
\begin{equation}
    z=e^{i\theta} = \cos(\theta) + i\sin(\theta). \label{eq:circz}
\end{equation}

We calculate our likelihood functions, $\mathcal{L}(\varphi_k)$, numerically using a grid approximation, by splitting the $2\pi$ support into $K=1024$ equally-sized bins such that $\theta_k = \theta_0+k/K,\ k\in\{0,1,2,...,K-1\}$ and calculating the value of the likelihood function for each bin. 

We combine two or more likelihood functions to find the likelihood function of the sum of some of the $\varphi_k(m)$ (with the same $m$) parameters by convolving them using fast Fourier transforms. This reduces the order of operations for the convolution from $\mathcal{O}(K^2)$ to $\mathcal{O}(K\log K)$ operations and can convolute as many likelihoods as we like in a single calculation providing further efficiency improvements. This produces likelihoods $\mathcal{L}(q\theta|\vec{n}), \ q\in\mathbb{N}^+$ from which we draw $\mathcal{L}(\theta|\vec{n})$. We do this for all of the $\varphi_k$ for which we have enough data then take the product of all the $\mathcal{L}(\theta|\vec{n})$ and normalise to get a final normalised likelihood function for the protocol. 

Drawn from the distance between two angles $1-\cos(\beta-\alpha)$ measure of circular dispersion~\cite{DirectionalStats} for some circular measurement data or distribution $\vec{\beta}$ around some point $\alpha$ is 

\begin{equation}
    D(\alpha) = \frac{1}{M} \sum_{m=1}^M \{1 - \cos(\beta_m-\alpha)\}.
\end{equation}

From a Bayesian perspective, the posterior distribution given by Bayes' rule,

\begin{equation}
    p(\theta|\vec{n},\alpha) \propto \mathcal{L}(\theta|\vec{n},\alpha)p(\theta|\alpha),
\end{equation}
is the probability distribution of the parameter $\theta$ given the measurement data $\vec{n}$ and a prior distribution $p(\theta|\alpha)$ where $\alpha$ is the prior distribution's hyperparameters (sometimes omitted when writing the posterior distribution alone). For this article we have used a uniform prior distribution $p(\theta|\alpha) = \frac{1}{2\pi} $ over an arbitrary $2\pi$ range. Therefore, the posterior distribution is equal to the normalised likelihood function. This broad prior with limited data creates broad posterior distributions which requires that we use allows us to make a circular analogue to the mean square error. We get this from the circular dispersion around the true parameter value $D(\theta)$ to the grid approximation of the likelihood function in equation~\ref{eq:circularMSE} to get $\lambda(\vec{n},\vec{\phi}) = \int D(\theta) d\mathcal{L}(\theta|\vec{n})$. When this is averaged over many sets of data for a set of input parameters it gives $\Lambda$ as a measure of information gain for those parameters.

While logical to combine all $\binom{B}{m}$ parameters $\varphi_k(m)$ that have the same $m$ to get $\binom{B-1}{m-1}$, like we do to calculate the Fisher information, it is advantageous both from a numerical efficiency standpoint and a limited data analysis perspective to take a more refined approach. An extreme example is when there is no data for one $\varphi_k(m)$ then it has a uniform distribution, like its prior and it's circular convolution with any other circular distribution is also uniform. This can render the information for all of the parameters we are attempting to combine with this no-data parameter useless for estimating $\theta$. 

As explained in section~\ref{sec:information_gain}, in limited data the number of results for each $\varphi_k(m)$ varies enough that it worth accounting for the variance of the combination of independent $J$ variables, $X_j$,

\begin{equation} \label{eq:VarianceSum}
    Var\left(\sum_{j=1}^J X_j \right) = \sum_{j=1}^J Var\left(X_j \right) = c \sum_{j=1}^J \frac{1}{n_j},
\end{equation}

for some constant c. We may use this principle as to create a figure of merit,

\begin{equation}
    n_{\text{effective}} = \left( \sum_{j=1}^J \frac{1}{n_j} \right)^{-1}
\end{equation}

for limited data analysis. If any $n_j=0$ then $n_{eff}=0.$ To optimise the information gain we make a list of all of the possible combinations of the $\varphi_k(m)$ with the same $m$ that sum to the smallest possible integer multiple of $\theta$ and search for the way of combining them that maximises the sum of the $n_{eff}$ and use those sets of combinations to estimate the likelihood function. The number of possible combinations increases rapidly with the number of Bobs, in particular when $m\approx N_B/2$. Therefore, to improve computation speed for larger numbers of Bobs (eg 16) when the number of combinations is very large we perform the optimisation for the $\varphi_k(m)$ with the largest $n_j$, record the best combinations, add the next largest to the remaining set and repeat until no non-zero data combinations are available.

\section{Optimisation algorithm} \label{sec:MCMC_optimisation}

We found the minimum dispersion for Alice given the security conditions and round count by searching through the possible protocol parameters. First, for $N_B=\{1,2,3,4,5,6\}$ Bobs and $N_R=\{100,500,2500\}$ rounds an evenly spaced $11\times 11$ grid of possible $P_S$ and $P_F$ in the range $[0,1]$, we performed the simulations of Alice's information gain for N rounds and Alice's information gain until the first time she is detected 16 times for 64 sets of B randomly chosen parameters (the same set for all simulations for $\{N_B,N_R\}$). We calculated $\mathcal{L}(\theta|\vec{n})$ for each set of results and took the average for every grid point to find $\Lambda_A$ and $\Lambda_E$ at that point in that grid.

Figs.~\ref{fig:SecurityE4differentBs} and~\ref{fig:SecurityS4differentBs} demonstrate the security is monotonic in both $P_F$ and $P_S$ for each $N_B$. The Fisher information is also monotonic in the opposite direction. However, as demonstrated in section~\ref{sec:information_gain} the limited data information gain is not necessarily monotonic. Due to this, we devised an optimisation algorithm where we find the positions in each grid direction where the security condition $\Lambda_E\geq 0.5$ is met and $\Lambda_A$ is minimised for each $P_S$. Then, we build a new, evenly spaced, grid with half the spacing of the previous grid out of those points and all of the points between them and repeated the previous step. We repeated the optimisation step 3 times and used the results of the simulations for the single grid point that minimised $\Lambda_A$ while $\Lambda_E\geq0.5$ for each $\{N_B,N_R\}$ to make the plots in section~\ref{sec:protocol}. 

 We created Fig.~\ref{fig:informationGainQuantumAdvantage} by taking the best optimised values $\{P_S,P_F\}$ for $N_B=\{1,2,3,4,5,6\}$ and calculating $\Lambda_A$ for $100$ values in the range $[1,2500]$ by using the same simulations for Alice and plotting them against similar simulations with $P_F=0$ and either $P_S=0$ or $P_S=1$.

\section{Fisher information of protocol} \label{sec:FisherInformation}

When calculating the Fisher information for the whole protocol relative to the number of rounds we use the probability of each $\varphi_k$ occurring. The following calculation will use $\mathcal{I}_m$ for the Fisher information, both classical and quantum when m parameters are measured together. First, considering the (quantum) Fisher information in a protocol using only separable initial states. Each Bob has a probability $P_M$ of performing a parameter estimation. Therefore, measurements of each of the $\phi_b$ will occur with a probability of M per round. Measuring each parameter separately and summing them to get an estimation for $\theta =  \sum_{b=1}^{N_B} \phi_b$ gives a (quantum) Fisher information

\begin{equation}
    \mathcal{I}(\theta|\text{separable only protocol}) = \frac{P_M \mathcal{I}_1 }{N_B}.
\end{equation}

If the protocol has a mix of states, the information gain due to separable states would be weighted by the probability of a separable initial state occurring

\begin{equation} \label{eq:FIfromS}
    \mathcal{I}(\theta,S) = \frac{ P_SP_M\mathcal{I}_1 }{N_B}.
\end{equation}

When considering entangled initial states, the probability of each sum of $m$ parameters $\varphi_{k}(m), k\in\{1,2,...,\binom{N_B}{m}\}$ being measured is equal. Therefore, it is practical to combine them to find the (quantum) Fisher information due to all of those parameters. They each have an equal probability

\begin{equation}
    P(m,k|B,F,E) = P_EP_M^mP_F^{N_B-m}
\end{equation}

of occurring. Their sum is

\begin{equation}
    \sum_{k=1}^{\binom{N_B}{m}} \varphi_k(m) = \binom{N_B-1}{m-1}\theta.
\end{equation}

This is evident by considering choosing only and all those combinations that contain an arbitrarily specific parameter there remain m-1 parameters to pick out of a remaining set of B-1. Therefore, $\binom{N_B-1}{m-1}$ combinations contain each parameter and so the sum of all of them contains the same multiple of that parameter. The (quantum) Fisher information for a single measurement follows the relationship

\begin{equation}
    \mathcal{I}(aX) = a^2\mathcal{I}(X),
\end{equation}

and (quantum) Fisher information due to an equally probable set of A states that sum to aX is 

\begin{equation}
    \mathcal{I}(aX) = a^2\mathcal{I}(X)/A.
\end{equation}

Therefore, the (quantum) Fisher information from the $\binom{N_B}{m}$ sums of m of the B parameters have a Fisher information

\begin{multline}
    \mathcal{I}(\theta|\text{entangled, m parameters in a round}) \\ =  \frac{\binom{N_B-1}{m-1}^2}{\binom{N_B}{m}} \mathcal{I}_m = \frac{m}{N_B}\binom{N_B-1}{m-1} \mathcal{I}_m.
\end{multline}

Combined with the occurrence probability of each individual state, this contributes 

\begin{equation} \label{eq:FIfromEm}
    \mathcal{I}(\theta,E,m) = P_EP_M^mP_F^{N_B-m} \frac{m}{N_B}\binom{N_B-1}{m-1}  \mathcal{I}_m.
\end{equation}

The (quantum) Fisher information combines additively so, for independent data X and Y

\begin{equation}
    \mathcal{I}_{X,Y}(\theta) = \mathcal{I}_{X}(\theta) + \mathcal{I}_{Y}(\theta).
\end{equation}

Therefore, the (quantum) Fisher information for the entire system relative to the number of rounds may be found by adding the information from equations \ref{eq:FIfromS} and \ref{eq:FIfromEm} for $m=1,2,...N_B$,

\begin{equation}
    \mathcal{I}_\text{total} = \frac{ P_SP_M\mathcal{I}_1 }{N_B} + \sum_{m=1}^{N_B} P_EP_M^mP_F^{N_B-m} \frac{m}{N_B}\binom{N_B-1}{m-1}  \mathcal{I}_m.
\end{equation}

The probability of each $\varphi_k$ occurring is dependent only on the number of parameters measured $m$,

\begin{equation} \label{eq:probK2}
    P_k(m)  = P_SP_M\Big|_{m=1} + P_EP_M^mP_F^{N_B-m}.
\end{equation}


\bibliography{main}

\end{document}